\documentclass[times,twocolumn,final,authoryear]{elsarticle}

\usepackage{jasr}
\usepackage{framed,multirow}

\usepackage{amssymb}
\usepackage{latexsym}
\usepackage{longtable}
\usepackage[switch]{lineno}

\usepackage{url}
\usepackage{xcolor}
\definecolor{newcolor}{rgb}{.8,.349,.1}
\usepackage{footnote}
\usepackage[citebordercolor=white]{hyperref}

\journal{Advances in Space Research}

\begin{document}

\verso{Given-name Surname \textit{etal}}

\begin{frontmatter}

\title{The flux ratio of the [N II]$\lambda\lambda$ 6548, 6583 \AA \ lines in sample of Active Galactic Nuclei Type 2}%

\author[1]{Ivan \snm{Doj\v cinovi\' c}\corref{cor1}}
\ead{ivan.dojcinovic@ff.bg.ac.rs}
\author[2]{Jelena \snm{Kova\v cevi\' c-Doj\v cinovi\' c}}
\ead{jkovacevic@aob.bg.ac.rs}
\author[2,3]{Luka \v C. \snm{Popovi\' c}}
\ead{lpopovic@aob.bg.ac.rs}

\address[1]{Faculty of Physics, University of Belgrade, Studentski Trg 12, Belgrade, Serbia}
\address[2]{Astronomical  Observatory,  Volgina  7, 11060  Belgrade, Serbia}
\address[3]{Faculty of Mathematics, University of Belgrade, Studentski Trg 16, Belgrade, Serbia}


\begin{abstract}
In spectra of the Active Galactic Nuclei (AGNs), the [N II]$\lambda\lambda$ 6548, 6583 \AA \ lines are commonly fitted using the fixed intensity ratio of these two lines ($R_{[N II]}=I_{6583}/I_{6548}$). However, the used values for fixed intensity ratio are slightly different through literature. There are several theoretical calculations of the transition probabilities which can be used for the line ratio estimation, but there are no experimental measurements of this ratio, since the [N II] lines are extremely weak in laboratory plasma.  
Therefore, the intensity ratio of [N II] lines can be measured only in the spectra of astrophysical objects. However, precise and systematic measurements have not be done so far, because of difficulties in measurement of the [N II] ratio in various spectra (overlapping with H$\alpha$, weak intensity of [N II], influence of the continuum noise and outflow contribution, etc.).
Here we present the measurements of the flux ratio of the [N II]$\lambda\lambda$ 6548, 6583 \AA \ emission lines for a sample of 250 Type 2 AGNs spectra taken form Sloan Digital Sky Survey (SDSS) data base. The spectra are chosen to have high signal-to-noise ratio and to [N II] and H$\alpha$ lines do not overlap. The obtained mean flux ratio from measurements is 3.049 $\pm$ 0.021. Our result is in agreement with theoretical result obtained by taking into account the relativistic corrections to the magnetic dipole operator. 
\end{abstract}


\end{frontmatter}


\section{Introduction}
 Nitrogen (N) is the seventh most abundant element  in the Universe and  it can be observed in the spectra of different astrophysical objects.
The forbidden emission lines [N II]$\lambda\lambda$ 6548, 6583 \AA \ were first time detected in optical spectra of gaseous planetary nebulae in our Galaxy \citep{Bowen1927,Bowen1928,Boyce1933}. \cite{Bowen1927} found that strong emission lines observed in spectra of gaseous nebulae, which have not been seen in any terrestrial source, originate by spontaneously emission from metastable level, which is possible only in extremely low density medium. In this way, he identified several forbidden emission lines, among them [N II]$\lambda\lambda$ 6548, 6583 \AA.

The development of astronomical observations has led to the frequent observation of these lines in spectra of various astrophysical objects. They  are among the most prominent narrow emission lines in spectra of the Active Galactic Nuclei - AGNs, HII regions/starburst galaxies, planetary nebulae, nova shells of gas and supernova remnants  \citep{Osterbrock2006}. 
In AGNs, they arise in low density ($<$10$^5$ cm$^{-3}$) photoionized gas, called Narrow Line Region (NLR), which has lower velocity dispersion (up to 500 km s$^{-1}$) comparing the velocity of the gas in the Broad Line Region (BLR), which is closer to the super massive black hole in the center of an AGN \citep[see][]{Osterbrock2006}. The [N II] lines could also arise in outflows, the large-scale phenomena, which are commonly seen in AGNs \citep{Woo2016, Kovacevic2022}.

 The [N II]$\lambda\lambda$ 6548, 6583 \AA \ lines have the similar term structure as [O III] $\lambda\lambda$5007, 4959 \AA, which are the most prominent narrow emission lines in low-density photoionized gas. Similarly as [O III] lines, [N II] lines originate from the same upper and slightly different lower energy level and have a negligible optical depth since the transitions are strongly forbidden. Therefore [N II]$\lambda$ 6548 \AA \ and [N II]$\lambda$ 6583 \AA \ lines may be scaled to exactly the same emission line profile, as it is shown for [O III] lines \citep{Dimitrijevic2007}.

Because of observational and physical circumstances, these two pairs of lines ([N II] and [O III]) are suitable for diagnostics of astrophysical objects \citep{Osterbrock2006}. They can be used for investigation of the gas kinematics in the Narrow Line Region of AGNs \citep{Eun2017} and AGN outflow dynamics \citep{Freitas2018}. The line intensity ratio of [N II] $\lambda$6583/H$\alpha$ is used in  BPT (Baldwin, Phillips, Terlevich) diagnostic diagram \citep{Baldwin1981} for classification of extragalactic objects. The BPT diagram consists of emission-line intensity ratios [N II]$\lambda$6583/H$\alpha$  versus [O III]$\lambda$5007/H$\beta$, supplemented by separation curves of \cite{Kewley2001} and \cite{Kauffmann2003}. Using this diagram one can identify the nature of the gas ionization source \citep{Veilleux1987, Smirnova2007}, and separate extragalactic objects to AGNs, starburst galaxies or LINERs (Low Ionization Nuclear Emission-line Region galaxies). In some cases, separation between starburst galaxies and AGN-type objects is made only by [N II]/H$\alpha$  ratio \citep{Bae2017}.  Also, the [N II]/H$\alpha$  ratio is commonly used as metallicity indicator \citep[see][]{Groves2006,  Martens2019}.

Since the H$\alpha$ line is positioned between the [N II]$\lambda$6548 \AA \ and [N II]$\lambda$ 6583 \AA \ lines and there are small wavelength differences between these lines, [N II] and H$\alpha$ lines are often blended in spectra of AGNs, which sometimes makes difficult to use  these lines in diagnostics purposes. To reduce the number of fitting parameters in decomposition of the complex [N II]+H$\alpha$ wavelength band it is necessary to fix the intensity ratio of [N II] $\lambda\lambda$6583, 6548 \AA \ lines ($R_{[N II]}=I_{6583}/I_{6548}$). Since these two  lines have the same upper level of the transition, it is expected that their intensity ratio is fixed in spectra, as it is  the case for the [O III]$\lambda\lambda$ 5007, 4959 \AA \ lines \citep{Dimitrijevic2007}. 

However, the used values for $R_{[N II]}$ are slightly different through literature and there is a need for accurate measurement of this ratio. The obtained theoretical values for [N II]$\lambda$6583.45 \AA/[N II]$\lambda$6548.05 \AA \  transition probability ratio ($A_{6583}/A_{6548}$) are between 2.93-3.07 (see Table \ref{T1}). \cite{Acker1989} measured the flux ratio of [N II] lines  for 267 planetary nebulae of our Galaxy. They found the line ratio of $R_{[N II]}$ = 2.92$\pm$0.32. However, there is lack of the systematic experimental measurements of [N II] lines from AGN spectra. These measurements are done only for several particular objects \citep[see][]{Nazarova1996, Dietrich2005}. 

The aim of this work is to do systematic and accurate measurements of [N II] lines ratio in large sample of AGNs Type 2 spectra and to compare the obtained value with various theoretical values. In Section \ref{Sec2} we give some basic properties of the N$^+$ ion and the review of the different theoretical results for [N II] lines ratio. In Section \ref{Sec4} we described the properties of the [N II] lines in AGN spectra and we gave some measured values of [N II] ratio for particular objects from literature. The procedure of the sample selection and spectral analysis are presented in Section  \ref{Sec5}. In Section \ref{Sec6} we give our result of 
[N II] ratio and compare it with theoretically obtained values, and in Section \ref{Sec7} we outline some conclusions.

\section{Theoretical values of the [N II]$\lambda$6548/[N II]$\lambda$6583 ratio }\label{Sec2}

 Ionization energy, necessary for ion N$^+$ forming, is 14.5 eV. Ion N$^+$ is carbon-like, as well as O$^{+2}$ ion, with electron configuration of ground state 2s$^2$2p$^2$. Allowed terms for p$^2$ equivalent electrons are $^3$P, $^1$D and $^1$S. Ground level in [N II] spectra is $^3$P$_0$. Another levels of $^3$P term are $^3$P$_1$ (with energy 0.00604 eV), and $^3$P$_2$ (0.01622 eV) \citep[data taken from the National Institute of Standards and Technology (NIST) database, see][]{Kramida2020}. First excited level is metastable $^1$D$_2$, with 1.89897 eV energy. The $^1$D state have low excitation potentials, high statistical weights and long lifetime of about 300 s \citep{Bowen1936}. 
 
There are three possible transitions between metastable $^1$D$_2$ and $^3$P terms ($^3$P$_0$, $^3$P$_1$, $^3$P$_2$). Two transitions, which are observed in astrophysical spectra are:

\begin{enumerate}

\item 2s$^2$2p$^2$ $^1$D$_2$ - 2s$^2$2p$^2$ $^3$P$_1$ ([N II]$\lambda$ 6548.05 \AA),

\item 2s$^2$2p$^2$ $^1$D$_2$ - 2s$^2$2p$^2$ $^3$P$_2$ ([N II]$\lambda$ 6583.45 \AA).


\end{enumerate}

The radiative transitions within $^1$D and $^3$P terms are forbidden in electric dipole radiation, because they all have the same parity, i.e. they violate $\Delta$S = 0 rule (intercombination transitions). Although they occur with small transition probabilities, the [N II]$\lambda$ 6548 \AA \ and [N II]$\lambda$ 6583 \AA \ lines are observed in optical spectra of different astronomical sources as strong emission lines, because of very low densities and a huge amount (volume) of photoionized gas where they arise. Their transition probabilities, calculated with different theoretical models, are given in the first and second column of Table \ref{T1}. 

The third possible transition, [N II] $\lambda$6527.23 \AA \ (2s$^2$2p$^2$ $^1$D$_2$ - 2s$^2$2p$^2$ $^3$P$_0$), is additionally forbidden by selection rule $\Delta$J = 0, $\pm$ 1 and therefore it has the smallest transition probability (10$^{-7}$ s$^{-1}$). That line is not visible in the spectra. 

The critical density of [N II] $^1$D$_2$ for collisional deexcitation is 6.6 $\cdot$ 10$^4$ cm$^{-3}$ \citep{Osterbrock2006}. In medium with higher electron densities, metastable excited level $^1$D$_2$ is depopulated by electron impact rather than radiation transition, and [N II]$\lambda\lambda$ 6548, 6583 \AA \  lines do not occur in spectrum.

\subsection{The comparison of the theoretical values }\label{Sec2.2}

Here we review some theoretical works which try to explain the  $^3$P$_1$ - $^1$D$_2$ and $^3$P$_2$ - $^1$D$_2$ transitions ([N II]$\lambda\lambda$ 6548, 6583 \AA). Since these singlet-triplet transitions are forbidden in electric dipole approximation, they are possible by the breakdown of Russell-Saunders coupling. The wave function of the $^1$D$_2$ level is a linear combination of $^1$D$_2$ and $^3$P$_2$ levels, so the transition to a triplet state can occur through this small part of the triplet in the singlet dominant wave function. Therefore, the mixing of different multiplicity states occurs through the spin-orbit interaction, which is very important for occurrence of the [N II] lines \citep{Condon1959}. While [N II]$\lambda\lambda$ 6548, 6583 \AA \ lines are essentially due to magnetic dipole radiation, the third line [N II]$\lambda$6527 \AA, which is not visible in spectra, is made due to electric quadrupole only, and its transition probability is of the order of 10$^{-7}$ s$^{-1}$ \citep{Kramida2020}. Contribution of the  electric quadrupole in  [N II]$\lambda$6548 and [N II]$\lambda$6583 lines intensity is about 0.1 per cent. One can see that the electric quadrupole radiation for these lines is of the order of 10$^{-6}$ s$^{-1}$ \citep{Galavis1997}.

\begin{table*}
\centering
\caption{In the first three columns are given theoretical results of transition probabilities of [N II]$\lambda$6583 \AA \ ($A_{6583}$) and [N II]$\lambda$6548 \AA \ ($A_{6548}$) lines and their transition probability ratio ($A_{6583}/A_{6548}$). In last column are given the references for the calculated data.
}\label{T1}
\begin{tabular}{|c|c|c|l|}
\hline
 
$A_{6583}$   {\footnotesize (s$^{-1}$)}    &  $A_{6548}$ {\footnotesize  (s$^{-1}$)}  &  $A_{6583}/A_{6548}$ &  References  \\
\hline

  --     &  --    & 3  & \cite{Stevenson1932}     \\

 2.4$\cdot$10$^{-3}$       & 0.81$\cdot$10$^{-3}$       & 2.96  & \cite{Condon1934}      \\

 2.2$\cdot$10$^{-3}$       & 0.75$\cdot$10$^{-3}$    & 2.93  & \cite{Pasternack1940}   \\

 2.99$\cdot$10$^{-3}$       & 1.01$\cdot$10$^{-3}$       & 2.96  &  \cite{Mendoza1983}     \\

 3.005$\cdot$10$^{-3}$    & 1.016$\cdot$10$^{-3}$     & 2.96 & \cite{Galavis1997} \\

 3.015$\cdot$10$^{-3}$     & 0.9819$\cdot$10$^{-3}$      & 3.07  & \cite{Storey2000}    \\

 2.91$\cdot$10$^{-3}$       & 0.984$\cdot$10$^{-3}$  & 2.96  & \cite{Tachiev2001}\footnotemark \\
\hline
\end{tabular}

\vspace{0.2 cm}
{\scriptsize [1] This reference is given in NIST database 2021 (\url{https://www.nist.gov/pml/atomic-spectra-database}, see \cite{Kramida2020}).
}
\end{table*}

Some theoretical results of [N II]$\lambda\lambda$ 6548, 6583 \AA \ radiative forbidden transition probabilities and their ratio are given in Table \ref{T1}.  \citet{Stevenson1932} took basic perturbation calculations and found that transition probability [N II] lines ratio ($A_{6583}/A_{6548}$) is exactly 3:1, with mean life of $^1$D$_2$  state of 3.1 minutes. Several authors obtained $A_{6583}/A_{6548}$ ratio of 2.96 \citep{Condon1934, Mendoza1983, Galavis1997, Tachiev2001}, while  \cite{Pasternack1940} obtained value of 2.93. 

\citet{Condon1934} considered the radiation field produced by an oscillating magnetic dipole.  Single-triplet transition occurs because of the magnetic spin-orbit interaction, where the levels $^1$D$_2$ and $^3$P$_2$ interact, corresponding to J = 2, and forming upper level with dominant single and small contribution of the triplet state. In order to calculate $A_{6583}/A_{6548}$ ratio, \cite{Pasternack1940} used Condon and Shortley theory of atomic spectra \citep[][first printed in 1935]{Condon1959}. \cite{Mendoza1983} included in calculations the relativistic interaction and electron correlation effects. \cite{Tachiev2001} calculated energy levels, lifetimes and transition data for carbon like emitters in the Breit-Pauli approximation.

\cite{Galavis1997}, calculated radiative rates within the IRON Project for the forbidden transitions within the ground configuration of atoms and ions in the carbon isoelectronic sequences. Atomic structure is calculated by SUPERSTRUCTURE code. The total radiative rate for a forbidden transition is taken to be the sum of the electric quadrupole (E2) and magnetic dipole (M1) contributions. The wave functions are calculated in two types of potential. Spectroscopic orbitals (1s, 2s and 2p) are calculated in a statistical Thomas-Fermi-Dirac model potential, while other electron radial wave functions are obtained in a scaled Coulomb potential. \cite{Galavis1997} also used the so-called Term Energy Corrections. \cite{Storey2000} used the same configuration interaction model as \cite{Galavis1997}, but with included Breit-Pauli  relativistic corrections to the 
magnetic dipole operator. Their relativistic corrections include all the one-body and two-body magnetic fine-structure interactions, spin-orbit, spin-spin etc. The electric quadrupole operator result is taken from \cite{Galavis1997}. Finally, \cite{Storey2000}  obtained the $A_{6583}/A_{6548}$ ratio of 3.07, that is slightly larger value than values calculated by other authors.

\section{The [N II]$\lambda\lambda$ 6548, 6583 \AA \ lines in AGN spectra}\label{Sec4}

 The velocity dispersion of the [N II] lines in Type 2 AGN spectra approximately follows one-to-one relationship with stellar velocity dispersion which indicates that these lines partly arise in the gravitationally bounded gas \citep[see][]{Eun2017, Kovacevic2022}. Therefore, they are commonly modelled with a single Gaussian, dominantly broaden by Doppler effect, which velocity dispersion (following stellar velocity dispersion) rarely exceed the $\sim$ 200 km s$^{-1}$ \citep{Kovacevic2022}. If an outflow is present in an AGN structure, it contributes to the narrow emission line profile in the line wings. In these spectra, [N II] emission lines are commonly modelled with double Gaussian model: one Gaussian which fits the core of the line (core component), and represents emission from gravitationally bounded gas, and a second, broader Gaussian  which fits the wings of the line (wing component), and represents the simplified model for an outflow emission \citep[see diverse theoretical models of outflow emission shape in][]{Bae2016}. The wing Gaussian is commonly shifted to the blue or to the red comparing the core component, representing emission from an approaching or receding outflow cones, while centered wing component probably indicates superposed emission from both \citep{Kovacevic2022}.

It is commonly assumed that [N II] $\lambda$6583 \AA \ and [N II] $\lambda$6548 \AA \ lines arise in the same emission region, so it is expected to both lines have the same widths and shifts of the core components \citep{Popovic2004}, and  the same widths and shifts of the wing components (if they exist).  

\cite{Kovacevic2022} analyzed the sample of 577 spectra typically classified as Type 2 AGNs, and found that in 
$\sim$40\% of the cases, the [N II] and H$\alpha$ lines are blended, mostly due to strong outflow contribution in wing components. On the other hand, in the spectra of AGNs Type 1 (also Type 1.8 and 1.9), the narrow [N II] and  H$\alpha$ emission lines are additionally overlapped with broad H$\alpha$ line, which originate from the BLR of an AGN. In some cases, broad H$\alpha$ line is so strong that narrow [N II] lines cannot be seen at all. Therefore, the lack of the systematic measurement of the [N II] flux ratio in a large sample of AGN spectra, could be partly explained by difficulties to precisely extract the profiles of the [N II]$\lambda\lambda$ 6548, 6583 \AA \ lines from AGN Type 1 spectra and from large amount of the AGN Type 2 spectra, and to measure their flux ratio.

In order to decompose complex [N II]+H$\alpha$ wavelength band in AGN Type 1/Type 2 spectra, and to reduce the  number of fitting parameters, different values  were used througth literature to fix the flux ratio of [N II] $\lambda$6583 and [N II] $\lambda$6548 lines. The most frequently,  this flux ratio is fixed as 3:1, without giving any particular reference \citep[see e.g.][]{Zakamska2003, Mullaney2013, Faisst2018, Fischer2019, Manzano-King2019, Marasco2020, Davies2020}. However, in some papers, the flux ratio has been fixed at the theoretically obtained value of transition probability ratio  2.96 \citep[see e.g.][]{Popovic2004, Greene2004, Reines2013, Woo2014, Kawasaki2017}. Also, other values have been used, as e.g. \cite{Zakamska2016}  used value of the ratio 2.94, \cite{Martens2019} value of 3.077, \cite{Nazarova1996} reported theoretical value of 2.88, etc.

Despite the fact that spectra of Seyfert galaxies and quasars have not been used to explicitly check the theoretical flux ratio of the [N II]$\lambda\lambda$ 6548, 6583 \AA \ lines, there are examples where such ratios were obtained as a by-product or could be derived from published results \citep{Nazarova1996, Cooke2000, Barcons2003, Dietrich2005}.  \cite{Nazarova1996} investigated the Seyfert 1.2  galaxy Mrk 79 with long-slit spectroscopy. Using their measured data it could be derived that  $R_{[N II]}$ in nucleus is  3.00, while in the different parts of the extended NLR, the ratios are in the range of 2.52-3.69, which might be caused by different signal-to-noise ratio (SNR) in long-slit spectra.

 \cite{Dietrich2005} measured the NLR emission line flux for 12 Narrow Line Seyfert 1 (NLSy1) galaxies. The derived  $R_{[N II]}$ for these objects are in the range of 2.92-3.26. 
From measurements of the line intensities given in \cite{Cooke2000} for Seyfert 2 galaxy NGC 3393, one may obtain $R_{[N II]}$ = 2.96, while in \cite{Barcons2003} for Seyfert 1.8/1.9  galaxy H1320+551, the measured [N II] flux ratio is 2.95. Note that [N II] lines are strongly blended by broad H$\alpha$ in \cite{Nazarova1996}, \cite{Dietrich2005} and \cite{Barcons2003} and by strong wing components in \cite{Cooke2000}, which makes the precise measurements of the [N II]$\lambda\lambda$ 6548, 6583 \AA \ line intensities very difficult.

On the other hand, \cite{Acker1989} performed the  measurements of the [N II] line intensities in large number of the  planetary nebulae spectra (267), but with low spectral resolution (1 nm), which affects the accuracy of measurements. The obtained result in \cite{Acker1989} is $R_{[N II]}$=2.92$\pm$ 0.32.

\begin{figure*}
\centering
\includegraphics[scale=0.9]{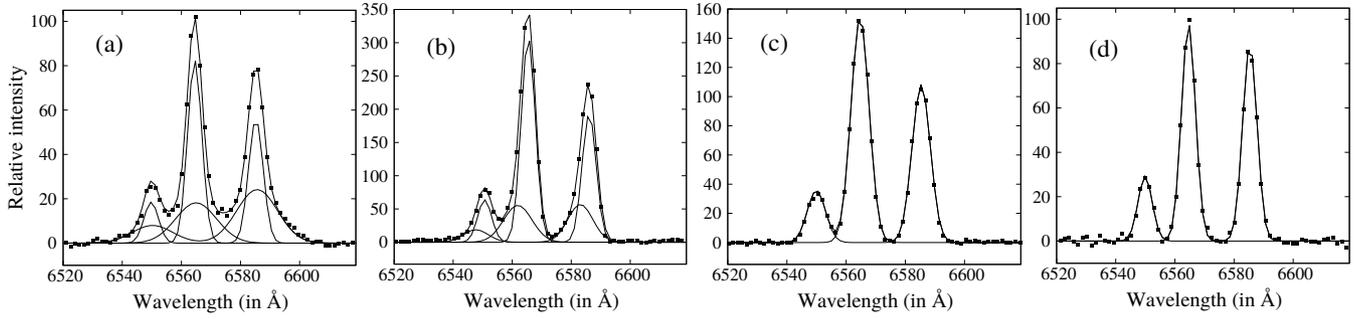}
\caption{Examples of [N II] + H$\alpha$ wavelength band in AGNs Type 2. (a) The spectrum with blended H$\alpha$ and N II lines, (b) the H$\alpha$ and [N II] lines are unblended, but fitted with double Gaussian model for each line, (c) both lines are fitted with single Gaussian, but they sligthly overlap at $\sim$ 6558 \AA \ and (d) the spectrum where H$\alpha$ and [N II] lines are fitted well with single Gaussian and they do not overlap. The spectra as in cases (a) and (b) are rejected from the sample.}
\label{fig1}
\end{figure*}

\section{The sample selection and fitting procedure}\label{Sec5}

 Diverse types of emission line galaxy spectra can be used for $R_{[N II]}$ measurements (Type 2 AGNs, starburst galaxies, LINERs, etc.). Disadvantage of the Type 2 AGN and LINERs spectra is that [N II] and H$\alpha$ lines could have multi-component shapes which overlap. On the other hand, in starburst galaxies, [N II] line intensity is smaller relative to H$\alpha$, comparing the same in the Type 2 AGNs. In this research, we use only Type 2 AGN spectra, since the idea is to find the $R_{[N II]}$ that can be used for decomposition of the Type 1 (Sy 1) AGNs with broad lines, where the $R_{[N II]}$ is usually 'apriori' assumed. We expect that NLR physical properties are the same (or quite similar) for all AGNs.

The spectra were  chosen from  Sloan Digital Sky Survey (SDSS) Data Release 14 (DR14) \citep{Abolfathi2018}, using Structural Query Language (SQL) with following requests: to be classified as Type 2 AGNs in SDSS spectral classification, median SNR over all good pixels in spectrum to be $>$ 20, and to have strong emission lines ([O III], H$\beta$, [S II], [N II] and H$\alpha$ EWs to be larger than 5 \AA). In this way, we obtained 588 spectra. The spectra were corrected for the Galactic reddening  and for the cosmological redshift. For correction of the Galactic reddening we used standard extinction law from \cite{ho1983} and  extinction coefficients from \cite{SF2011}. We applied the spectral principal component analysis (SPCA) in order to decompose the spectra to the host-galaxy and AGN contribution \citep[for details of the SPCA procedure see][]{Kovacevic2022}. After we obtained the host-galaxy spectra (with masked emission lines), we subtracted them from the observed spectra. In this way we obtained the AGN contribution which is not affected with stellar absorption features.

From the initial sample, through several steps, we selected the subsample  which is optimal for precise measurements of the [N II] flux ratio. It is the subsample where H$\alpha$ and [N II] lines could be fitted with a single-Gaussian model, i.e. they do not have wing components. Presence of the wing components in these lines import additional degrees of freedom in fitting process, and cause overlapping H$\alpha$+[N II].

First, we rejected all spectra where H$\alpha$ and [N II] lines are strongly blended, as shown in Figure \ref{fig1}a. We kept in sample only the spectra in which the  emission line flux at $\sim$ 6575 \AA \ (F$_{6575}$), which is minimum between H$\alpha$ and [N II] 6583 \AA \ line, is close to the continuum level (F$_{6575}$/N $<$ 3). This criterium left 346 spectra in sample. Then, we fitted simultaneously the continuum level within the range $\lambda\lambda$6510-6630 \AA \ with a linear function, and [N II]+H$\alpha$ lines with a single-Gaussian or a double-Gaussian model. For fitting process we used  nonlinear least-squares (NLLS) Marquardt-Levenberg algorithm. The H$\alpha$ line is fitted with all free parameters. The widths and shifts of Gaussians which fit [N II]$\lambda\lambda$ 6548, 6583 \AA \ lines were forced to be the same \citep{Popovic2004}, while their intensities were left to be the free parameters.

To distinguish whether a single or a double-Gaussian model should be applied on [N II]+H$\alpha$, we used the criterium of the extra-sum-of-squares F-test \citep{Lupton1993}. This test compares the improvement of sum-of-squares with the more complicated model  vs. the loss of degrees of freedom. The test is done for single-Gaussian versus double-Gaussian model for [N II] and H$\alpha$ lines in each spectrum. Double-Gaussian model was adopted if P-value $<$ 0.05, having assumed as null hypothesis that there is no significant improving of the fit by applying the model which includes more parameters (double-Gaussian), with respect to the model with smaller number of parameters (single-Gaussian).  Afterwards, we excluded from the sample all objects where [N II] and H$\alpha$ lines cannot be fitted well with single-Gaussian model (Figure \ref{fig1}b), i.e. where F-test P-value is $<$ 0.05. Finally, our sample contains 250 spectra of AGNs Type 2, where [N II] and H$\alpha$ lines can be fitted well with single Gaussian function for each line (Figures \ref{fig1}c,d). We found that in 75\% of the final sample [N II]$\lambda$6548 \AA \ and H$\alpha$ lines are slightly overlapped (as shown in Figure \ref{fig1}c), while in 25\% of the sample all three lines are well resolved (as in Figure \ref{fig1}d).

\section{Results and discussion}\label{Sec6}
\begin{figure}
\centering
\includegraphics[scale=0.28]{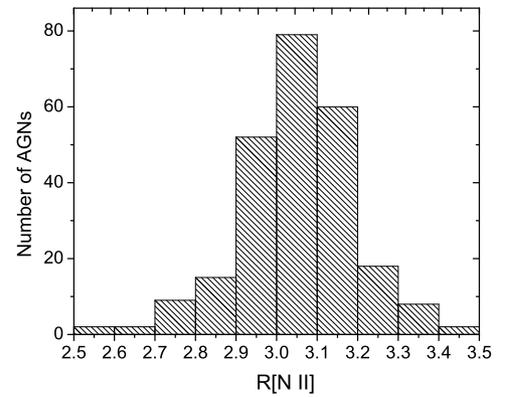}

\caption{Histogram of distribution of the [N II]$\lambda\lambda$ 6548, 6583 \AA \ lines ratio obtained from the best fit.}
\label{fig2}
\end{figure}

After we obtained the intensities of the [N II] lines from the best fit, we calculated their intensity ratio $R_{[N II]}$ for each object. The histogram of distribution of obtained $R_{[N II]}$  in sample of 250 AGNs is shown in Figure \ref{fig2}.

\subsection{The uncertainty analysis}\label{Sec6.1} 

\begin{figure}[t]
\centering
\includegraphics[scale=0.35]{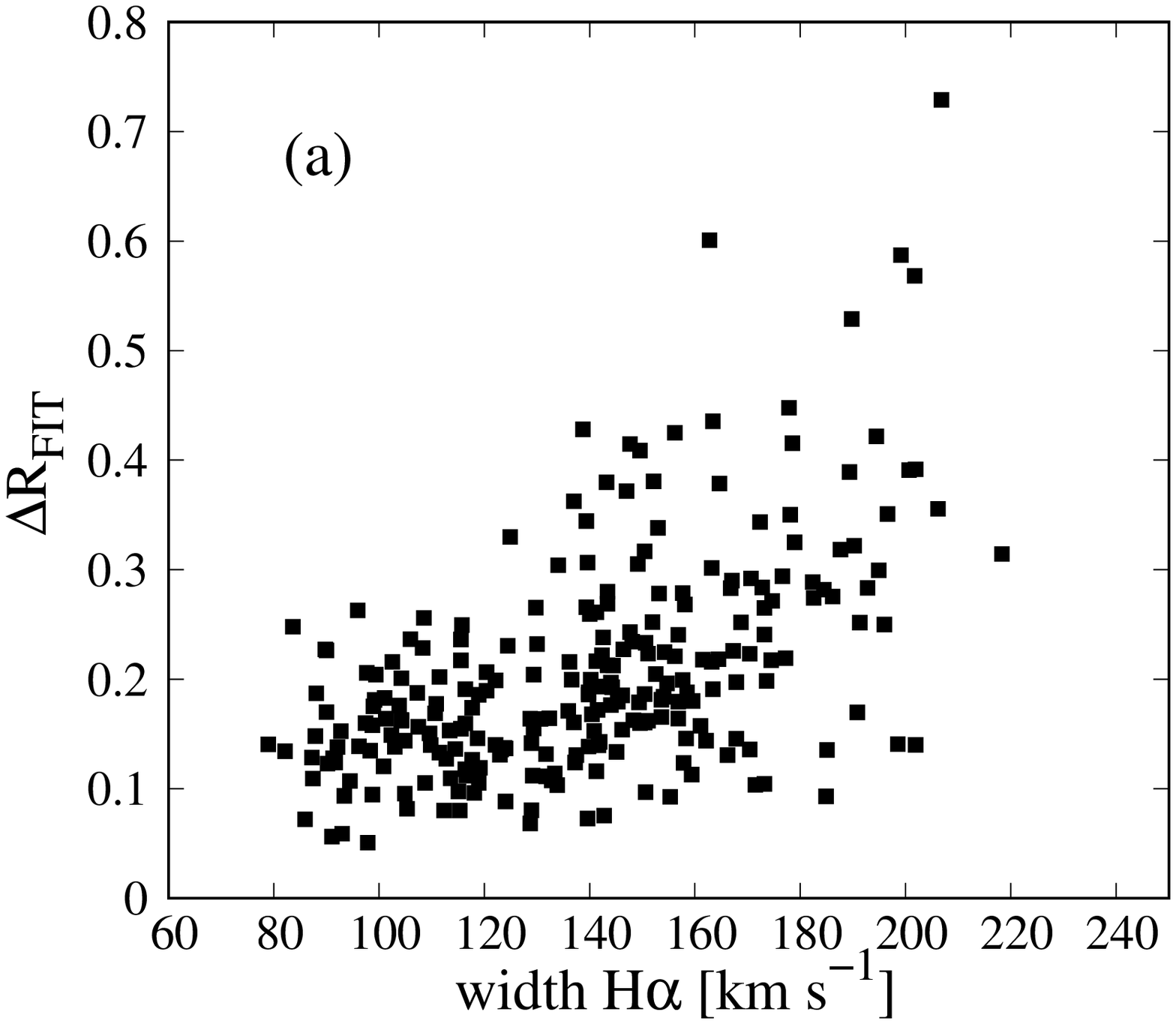}
\includegraphics[scale=0.35]{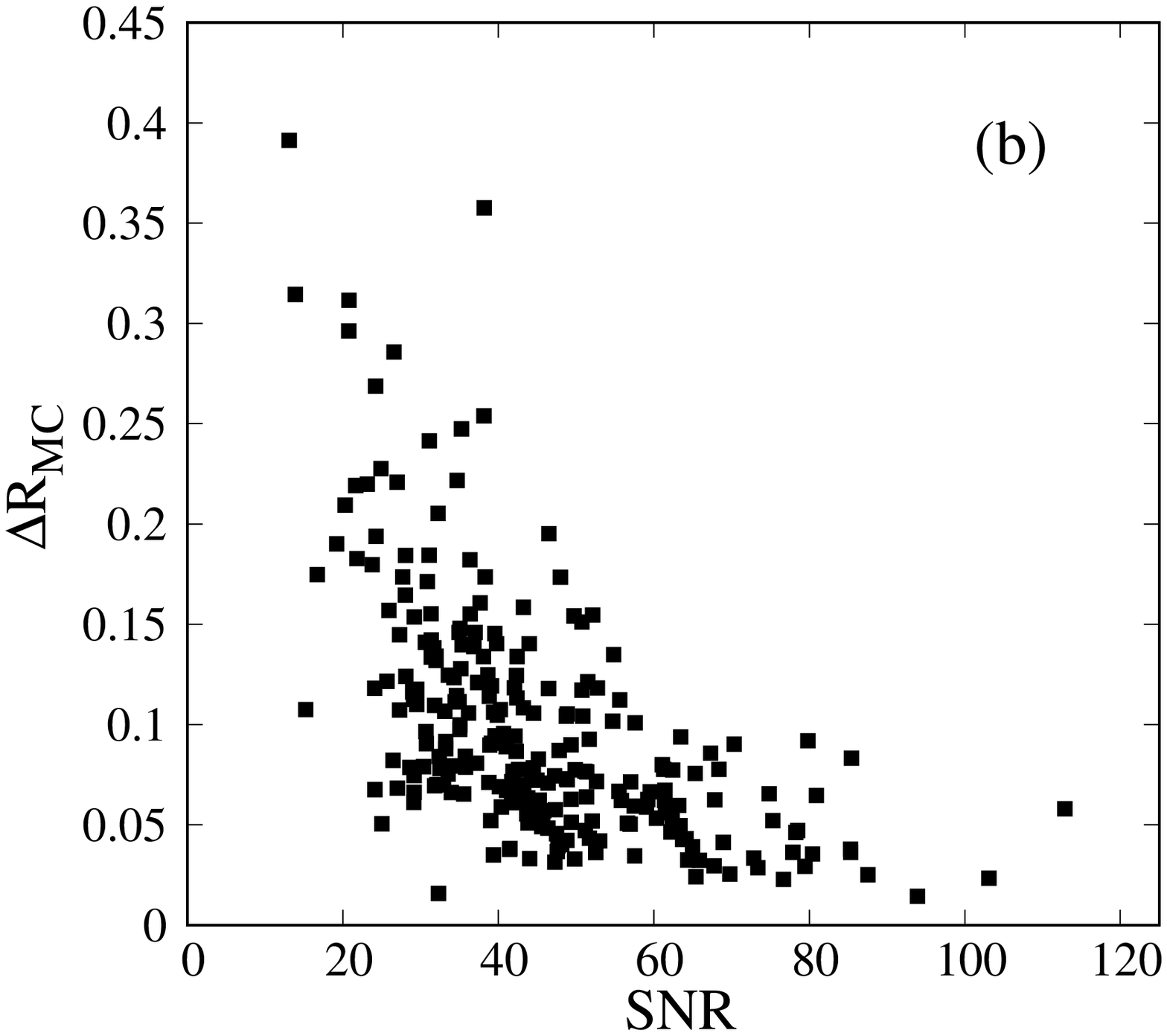}
\caption{Relationships of the estimated errors with spectral parameters: (a) the uncertainty obtained from Marquardt-Levenberg algorithm ($\Delta R_{FIT}$) vs.  H$\alpha$ velocity dispersion  and (b)  the uncertainty obtained by Monte Carlo method ($\Delta R_{MC}$) vs. SNR.}
\label{fig3}
\end{figure}

The  nonlinear least-square fitting method gives asymptotic standard errors for obtained fitting parameters. The standard errors of the [N II]$\lambda$6548 \AA \ and [N II]$\lambda$6583 \AA \ intensity parameters were used to estimate the  uncertainty of the [N II] line ratio caused by fitting procedure ($\Delta R_{FIT}$).  We found that the $\Delta R_{FIT}$ shows strong correlation with the widths of the H$\alpha$ and [N II] lines. As widths of these lines increase, the lines more overlap (as in Figure \ref{fig2}c), which makes larger uncertainty in determination of the [N II]$\lambda$6548 \AA \ intensity parameter. The correlation is shown in Figure \ref{fig3}a (Spearman coefficient of correlation is r = 0.51, and P-value = 0). In order to examine the influence of the noise to the results, we measured SNR in continuum near [N II]+H$\alpha$ ($\lambda\lambda$6610-6630 \AA) as the ratio of mean value of the flux continuum to standard deviation of the flux in the same range.
We found that $\Delta R_{FIT}$  is not strongly affected by the noise, since we found only negative trend between $\Delta R_{FIT}$ and SNR measured near [N II]+H$\alpha$.

 However, the role of the noise in [N II] intensity measurements is important, since it could affect significantly the shape of the smaller [N II]$\lambda$6548 \AA \ line, and also it contributes to the uncertainty of the continuum subtraction. Even a small difference in the determined continuum level and the slope due to noise could affect measured line intensities. Therefore, in order to additionally estimate the uncertainty in result due to noise, we applied Monte Carlo method for each spectrum. 
 First, we constructed the spectrum model for every object, using obtained values from the fit of emission lines and continuum level. We made 100 mock spectra for each source by adding the random noise to model spectra. The random noise is limited to have the same $\sigma$ of noise as it is measured in observed spectrum at $\lambda\lambda$6610-6630 \AA, and to have Gaussian distribution. Similarly as for observed spectra, we simultaneously fitted the continuum and emission lines in  mock spectra. 
After we obtained the fitting parameters of the mock spectra, we took the 1$\sigma$ dispersion of the parameters as the parameter uncertainty and we estimated  uncertainty of the [N II]$\lambda$6548/[N II]$\lambda$6583 ratio ($\Delta R_{MC}$). This procedure is repeated for each object. We tested if there is any correlation between $\Delta R_{MC}$ versus continuum/line properties and SNR. We found that $\Delta R_{MC}$ do not correlate with any continuum and line properties, contrary to $\Delta R_{FIT}$ which grows with larger line widths. On the other hand, $\Delta R_{MC}$ is strongly correlated with SNR as shown in Figure \ref{fig3}b (r = -0.64, and P-value = 0).

Since $\Delta R_{FIT}$ mainly represents uncertainty of [N II] ratio caused by line overlapping and $\Delta R_{MC}$ uncertainty of due to noise, we adopted sum of these two error estimates to be the absolute error for each measured line ratio ($\Delta R$ = $\Delta R_{FIT}$+$\Delta R_{MC}$). The weights of measurements are calculated as reciprocal square of the corresponding $\Delta R$, and the final value of the [N II] flux ratio is obtained as the weighted mean of the sample measurements. The uncertainty of the result is estimated  as error in the weigthed mean as given in \citet{Bevington}.
Finally, the obtained result for total sample is  $R_{[N II]}$ = 3.049 $\pm$ 0.021.

\subsection{The comparison with theoretical results}\label{Sec6.2}

Following the relationship $I \sim n_j \cdot A \cdot h \nu$, where $I$ is line intensity, $n_j$ the population of the level $j$, $A$ spontaneous transition probability and $h \nu$ energy of the transition, the theoretically expected line intensity ratio of [N II] lines ($I_{6583}/I_{6548}$) could be calculated using obtained transition probability ratio ($A_{6583}/A_{6548}$) as:

\begin{equation}\label{E1}
\frac{I_{6583}}{I_{6548}} = \frac{A_{6583}}{A_{6548}}\cdot \frac{6548 \AA}{6583 \AA}
\end{equation}

\begin{figure}[t]
\centering
\includegraphics[scale=0.35]{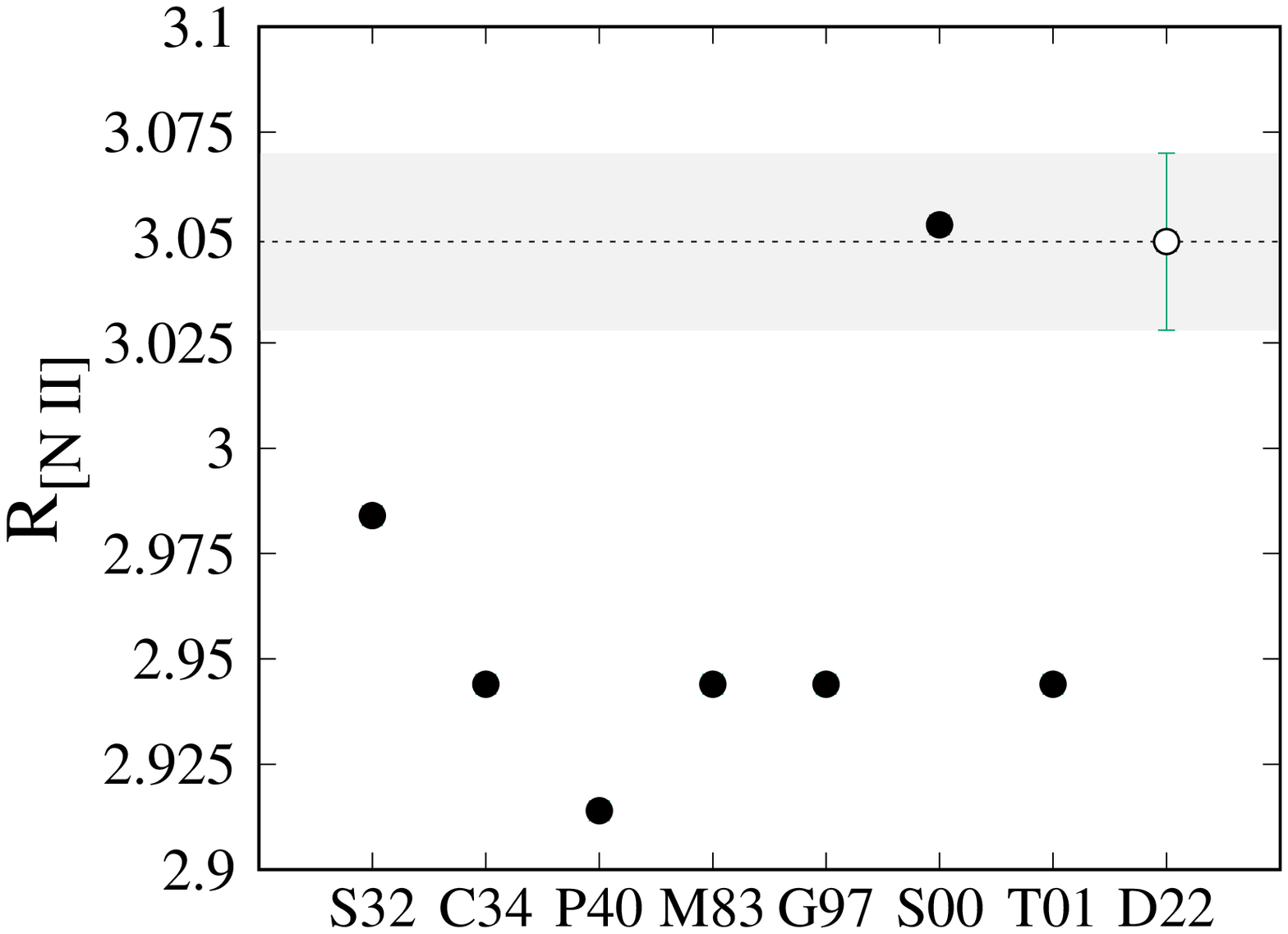}

\caption{Comparison of the $R_{[N II]}$ measured in this work (white dot) with theoretical results for [N II] intensity ratio (black dots). X -axis: S32 - \cite{Stevenson1932}, C34 - \cite{Condon1934}, P40 - \cite{Pasternack1940},
M83 - \cite{Mendoza1983}, G97 - \cite{Galavis1997}, S00 - \cite{Storey2000}, T01 - \cite{Tachiev2001} and D22 - this work. The  errorbar of our result is shaded with grey color. }
\label{fig4}
\end{figure}

 We applied Eq.~(\ref{E1}) to transition probabilities given in Table \ref{T1} and we compared various theoretically expected ratios with our result for $R_{NII}$ as shown in Figure \ref{fig4}. The theoretical calculation of \cite{Storey2000}, who took into account the relativistic corrections to the magnetic dipole operator, gives for $A_{6583}/A_{6548}$ = 3.07 (see Table \ref{T1}). This means that the expected line intensity ratio, following Eq.~(\ref{E1}), is $I_{6583}/I_{6548}$ = 3.05.
We concluded that our result of 3.049 $\pm$ 0.021 is in very good agreement with this theoretical result,  while theoretical results which do not include relativistic corrections are  smaller than our measured value, even within errorbar (see Figure \ref{fig4}). 

It is found that the observed line intensity ratio for [O III] $\lambda\lambda$5007, 4959 \AA \ lines is also in a good agreement with the theoretical result which include relativistic corrections obtained by \cite{Storey2000}.  \cite{Dimitrijevic2007} measured the flux ratio of the [O III]$\lambda\lambda$ 5007, 4959 \AA \ lines in a sample of the Type 1 AGN spectra and obtained the value of 2.993 $\pm$ 0.014, which is in accordance with theoretical value of 2.98 obtained in \cite{Storey2000}. It is interesting that this magnetic dipole operator relativistic correction effect becomes more significant for the near neutral ions. Namely, the magnetic dipole contribution is corrected about 4 \% for [N II] and about 3 \% for [O III]. Our observational result supports the introduction of theoretical relativistic corrections to the magnetic dipole operator in the calculation of the corresponding line intensity ratio.

\section{Conclusions}\label{Sec7}

In order to check the theoretical value of the intensity ratio of the [N II]$\lambda\lambda$ 6548, 6583 \AA \ lines, we measured the corresponding flux ratio in a sample of 250 Type 2 AGNs with a high-S/N spectra taken from the SDSS data base. We compared our  result with various theoretical results from existing literature, obtained with different approximations.
On the basis of our investigation, we give the following conclusions:
\begin{enumerate}
\item The intensity ratio of $I_{6583}/I_{6548}$ = 3.049 $\pm$ 0.021, obtained from our measurements, is in very good agreement with the theoretical improvement which includes relativistic corrections, obtained by  \cite{Storey2000}, who derived an line intensity ratio of 3.05. 
\item We found that the uncertainty of [N II] line ratio is dominantly caused by line overlapping (H$\alpha$ line is positioned between the [N II]$\lambda\lambda$ 6548, 6583 \AA \ lines) and uncertainty due to noise which affects the continuum subtraction. 
\item Despite the fact that the [N II] lines in spectra of AGN may have complex line profiles, often blended with H$\alpha$, they can be used to check sophisticated theoretical calculations by applying adequate sample selection and choosing the high quality spectra.

\end{enumerate}

 We may conclude that relativistic corrections to the magnetic dipole operator for $^1$D$_2$ - $^3$P$_1$ and $^1$D$_2$ - $^3$P$_2$ transitions in the carbon sequence are appropriate and must be taken into account to obtain the correct flux ratio of the emission lines.

\section{Acknowledgments}

This work is supported by the Ministry of Education, Science and Technological Development of Serbia (451-03-9/2021-14/20016 and 451-03-68/2020-14/200002).

Funding for the Sloan Digital Sky 
Survey IV has been provided by the 
Alfred P. Sloan Foundation, the U.S. 
Department of Energy Office of 
Science, and the Participating 
Institutions. 

SDSS-IV acknowledges support and 
resources from the Center for High 
Performance Computing  at the 
University of Utah. The SDSS 
website is www.sdss.org.

SDSS-IV is managed by the 
Astrophysical Research Consortium 
for the Participating Institutions 
of the SDSS Collaboration including 
the Brazilian Participation Group, 
the Carnegie Institution for Science, 
Carnegie Mellon University, Center for 
Astrophysics | Harvard \& 
Smithsonian, the Chilean Participation 
Group, the French Participation Group, 
Instituto de Astrof\'isica de 
Canarias, The Johns Hopkins 
University, Kavli Institute for the 
Physics and Mathematics of the 
Universe (IPMU) / University of 
Tokyo, the Korean Participation Group, 
Lawrence Berkeley National Laboratory, 
Leibniz Institut f\"ur Astrophysik 
Potsdam (AIP),  Max-Planck-Institut 
f\"ur Astronomie (MPIA Heidelberg), 
Max-Planck-Institut f\"ur 
Astrophysik (MPA Garching), 
Max-Planck-Institut f\"ur 
Extraterrestrische Physik (MPE), 
National Astronomical Observatories of 
China, New Mexico State University, 
New York University, University of 
Notre Dame, Observat\'ario 
Nacional / MCTI, The Ohio State 
University, Pennsylvania State 
University, Shanghai 
Astronomical Observatory, United 
Kingdom Participation Group, 
Universidad Nacional Aut\'onoma 
de M\'exico, University of Arizona, 
University of Colorado Boulder, 
University of Oxford, University of 
Portsmouth, University of Utah, 
University of Virginia, University 
of Washington, University of 
Wisconsin, Vanderbilt University, 
and Yale University.

\end{document}